\documentclass[aps,prl,twocolumn,nofootinbib,groupedaddress]{revtex4}
\usepackage[dvips]{graphicx}
\usepackage{amsmath,amssymb,graphicx,subfigure}

\newcommand{\be}{\begin{eqnarray}}
\newcommand{\ee}{\end{eqnarray}}

\def\slashchar#1{\setbox0=\hbox{$#1$}           
   \dimen0=\wd0                                 
   \setbox1=\hbox{/} \dimen1=\wd1               
   \ifdim\dimen0>\dimen1                        
      \rlap{\hbox to \dimen0{\hfil/\hfil}}      
      #1                                        
   \else                                        
      \rlap{\hbox to \dimen1{\hfil$#1$\hfil}}   

      /                                         
   \fi}                                         %
\newcommand{\ov}[1]{\overline{#1}}

\begin{document}
\title{Strong CP Violation in External Magnetic Fields}

\author{R. Millo$^{1}$ and P. Faccioli$^{1,2}$}
\affiliation{$^1$~Dipartimento di Fisica, Universit\`a degli Studi di Trento,\\ Via Sommarive 15, Povo (Trento) 38050 Italy. \\
$^2$~ I.N.F.N., Gruppo Collegato di Trento, Via Sommarive 15, Povo (Trento),38050 Italy.}

\email{faccioli@science.unitn.it}
\begin{abstract}
We study the response of the QCD vacuum to an external magnetic field, in the presence of strong CP violation. Using chiral perturbation theory and large $N_c$ expansion, we  show that the external field would polarize quantum fluctuations and induce an electric dipole moment of the vacuum, along the direction of the magnetic field. We estimate the magnitude of this effect in different physical scenarios.
In particular, we find that the polarization induced by the magnetic field of a magnetar  could accelerate electric charges up to energies of the order $\sim \theta \times~10^3~$TeV. 
We also suggest a connection with the possible existence of "hot-spots" on the surface of neutron stars.
\end{abstract}  
\maketitle

The request of a gauge-invariant definition of the vacuum gives raise to  the so-called $\theta$-terms in the QCD Lagrangian:
\be
S_\theta=  \theta \frac{1}{32 \pi^2}\int d^4x ~G_{\mu \nu} \tilde{G}^{\mu \nu},
\label{thetaterm}
\ee
where $\theta$ includes also a contribution from the weak sector, i.e.  $\theta= \theta_0 + \textrm{argdet}[M]$, where $M$ is the quark complex, non-hermitian mass matrix. 

The interaction term in Eq.~(\ref{thetaterm}) is a source of  CP-violation in the non-perturbative strong sector of the Standard Model~\cite{strongCP}.
At the moment, the most constraining bound $\theta\lesssim~3\times~10^{-10}$ comes from the measurement of the neutron electric dipole moment. In the context of the search for new physics, it is very important to quantify the amount of CP violation which has to be attributed to the Standard Model, and in particular to QCD. This motivates further research to provide better estimates of~$\theta$.

The main idea of the present work is to consider a CP-violating process in which the smallness of $\theta$ is compensated by the coupling to some other very large scale. 
To this end, we analyze the CP-odd response of the $\theta$-vacuum to a uniform external magnetic field and we show that the vacuum  develops an induced electric dipole moment along the direction of the external magnetic field. This effect vanishes for $\theta\to 0$ and is qualitatively different to the ordinary (i.e. CP-even) vacuum polarization,  which always occurs in the direction parallel to the external {\it electric} field. 

At zero temperature, the energy which can be delivered via such a mechanism to a charged particle in a magnetic domain with field strength $B$ extending for a distance $L$ is found to be proportional to $\theta B^3 L$. Finite temperature corrections scale like $\theta B L T^2$, for small $T$. 
Hence, it is natural to consider three different scenarios:  (i) a configuration in which the field is coherent over extremely large distance scales
(ii) a configuration in which the region permeated by the field is  hot and (iii) a configuration in which the external field is extremely intense.

In Nature, the last scenario is realized in the vicinity compact objects such as e.g. magnetars, where the magnetic field can reach strengths as high as $10^{15}\div10^{16}$~G~\cite{magnetar}. 
The first scenario is realized in regions of the observed Universe permeated by the so-called Large-Scale Magnetic Fields (LSMF). These are $\mu$G fields with correlation lengths as large as the size of galaxy clusters,~$\sim~10^2~\textrm{kpc}$~\cite{early}. 
According to the so-called primordial hypothesis, such fields are the result of the evolution of ``seed'' fields, which were formed in the early stages of the Big Bang,  when the temperature was large and therefore the second scenario may apply (for a review see e.g.\cite{cosmoBrev}). 
  

The starting point of our discussion is to express the electric-dipole density distribution in the vacuum in terms of a QCD matrix element:
\be
{\bf p}(t)&=& \frac{1}{V}\int d^3{\bf x}~{\bf x}\,\langle \theta |e J^{e/m}_0({\bf x}, t)|\theta\rangle_{A_\mu},
\label{edm0}
\ee
where $|\theta \rangle_{A_\mu}$ represents the $\theta$-vacuum state in the presence of the external field  $A_\mu$ and
$e J_{\mu}^{e/m}= e \sum_f Q_f \overline{q}_f \gamma_\mu q_f$, is the electro-magnetic current operator.
We  stress the fact that, in the absence of CP violation, an external electro-magnetic field cannot induce {\it electric} polarization along the direction of the ${\bf B}$ vector, hence ${\bf p}\cdot {\bf B}=0$.
 
In order to systematically account for the non-perturbative QCD dynamics in (\ref{edm0}), we adopt a chiral effective field theory description to $\mathcal{O}(p^4)$, with $2$ degenerate flavors, in which topological effects are accounted to leading-order in the $1/N_c$ expansion. Our generating functional is therefore
\be
Z_{EFT}[a_\mu,\theta] = \int \mathcal{D}U e^{i\int d^4 x \mathcal{L}_{EFT}[a_\mu,\theta]},
\label{Z}
\ee
where the effective Lagrangian is
\be
\mathcal{L}_{EFT}[a_\mu, \theta]=\mathcal{L}_{\chi pt}^{(2)}[a_\mu]+\mathcal{L}_{\chi pt}^{(4)}[a_\mu]+\mathcal{L}_{anom.}\label{L}[\theta].
\ee
$\mathcal{L}_{\chi pt}^{(2)}[a_\mu]$ and $\mathcal{L}_{\chi pt}^{(4)}[a_\mu]$ are the $\mathcal{O}(p^2)$ and $\mathcal{O}(p^4)$ chiral pertrubation theory Lagrangians respectively, including the mass term  and a gauge-invariant coupling to the  vector potential source term,  $a_\mu$. We recall that the $\mathcal{L}_{\chi pt}^{(4)}[a_\mu]$ term contains also the anomalous electro-magnetic Wess-Zumino coupling~\cite{WZ}.  

$\mathcal{L}_{anom.}[\theta]$ is a term which accounts both for the axial anomaly and for strong CP violation, to leading order in the $1/N_c$ expansion~\cite{DV}. After expanding the fields around their value in the vacuum, the effective Lagrangian can be written as~\cite{pich,otha}: 
\be
\mathcal{L}_{anom.}[\overline {\theta}]=  - \frac{ f_\pi^2  a}{ 4 N_c} 
\left[
\overline{\theta}^2- \frac{1}{4} \left( \log\left[\frac{\det U}{\det U^\dagger}\right]\right)^2\right.\nonumber\\
\left. + ~i \overline{\theta} \left( \textrm{Tr}(U-U^\dagger)- \log\left[\frac{\det U}{\det U^\dagger}\right]\right)
\right],
\ee
where $\frac{a}{N_c}$ is identified with the mass of the iso-singlet  pseudo-scalar meson (which for $N_f=2$ we shall denote with $\eta$) and $\frac{a}{N_c}\overline{\theta}\simeq \frac{1}{2}~m_\pi^2 \theta$.

\begin{figure}[t!]
\vspace{0.2 cm}
\includegraphics[width=0.18\textwidth]{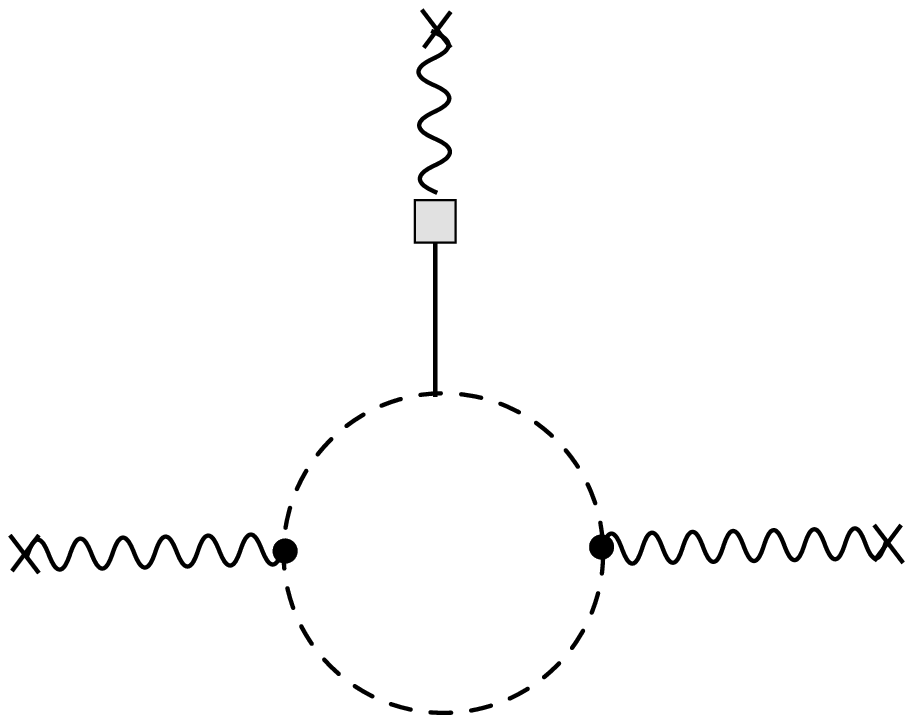}\hspace{.2cm}
\includegraphics[width=0.12\textwidth]{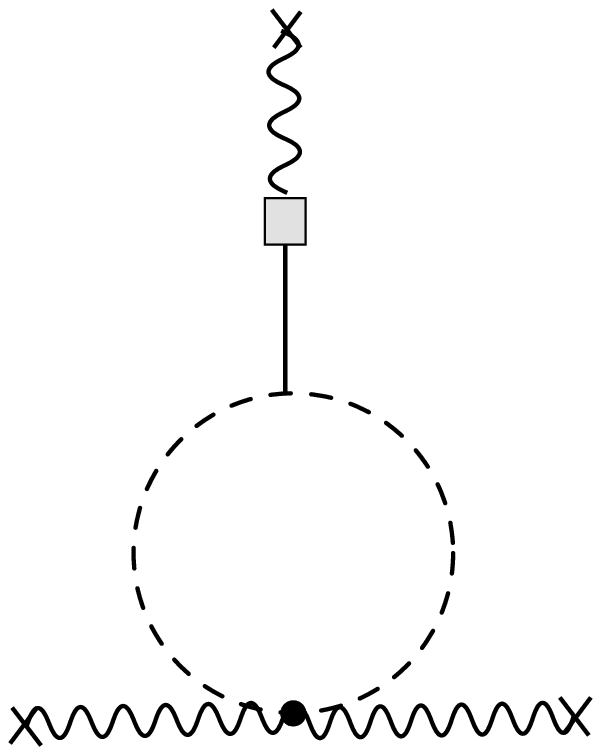}
\caption{Leading graphs in our estimate of the induced  vacuum electric dipole moment, at $T=0$. The dashed lines are pion propagators, the solid line are $\eta$ propagators and the gray square  denotes the Wess-Zumino electric charge density operator, Eq.~(\ref{WZe2}).}
\label{1Loop}
\end{figure}
In chiral perturbation theory, the matrix element of the charge operator in (\ref{edm0}) 
can be computed by functionally differentiating  the generating functional with respect to the external vector-potential source $a_\mu(x)$:
\be
\langle\theta| J_\mu(x)|\theta\rangle_{A_\mu} \simeq \left(\frac{\delta}{\delta i e a_\mu(x)} \log \mathcal{Z}_{EFT}\right)_{a_\mu=A_\mu}.
\label{edm2}
\ee

A useful topological property of the  CP violating diagrams contributing to the matrix element (\ref{edm2}) is revealed, when the relevant operators are expanded in powers of the meson fields.
The expansion of the electro-magnetic current operator 
\be
J_\mu(x) = \frac{\delta}{i e \delta a_\mu(x)} \int d^4 z  {\mathcal L}_{EFT}[a_\mu]
\ee
 contains terms with both odd and even powers of the fields. In particular, the terms coming from functionally differentiating $\mathcal{L}_{\chi PT}^{(2)}$ and the non-anomalous pieces of $\mathcal{L}_{\chi PT}^{(4)}$ display {\it even} powers of meson field operators,  while those coming from the anomalous Wess-Zumino term in $\mathcal{L}_{\chi PT}^{(4)}$ display {\it odd} powers of meson field operators.  On the other hand, the  CP-violating interaction $\mathcal{L}_{anom.}$ leads to a three-meson field vertex.
As a consequence, all CP-violating diagrams contributing to (\ref{edm2}) must contain the Wess-Zumino piece of the electro-magnetic current operator. This condition holds to any chiral orders, since it simply follows  from the requirement that propagator lines in the diagrams have to connect in a specific way in order to form closed loops. It is a manifestation of the fact that strong CP-violation is a purely quantum, anomaly-mediated process. 

In particular, the lowest-order diagrams (in the combined  chiral and $1/N_c$ counting and for small $\theta$) are those reported in Fig.\ref{1Loop}, and contain the Wess-Zumino electro-magnetic operator,
\be
e J^\mu_{WZ}= \frac{e^2 N_c}{48 \pi^2 f_\pi} ~\epsilon^{\mu \nu \alpha \beta  }  F_{\nu \alpha}~ \partial_\beta~ \left( \frac{5}{3} 
  \eta +  \pi_0\right), 
\label{WZe2}
\ee
along with one CP-odd three-meson interaction and with the non-anomalous  $\mathcal{O}(p^2)$ electro-magnetic vertexes.

In the physical scenarios we are presently interested in, the external field can be considered  static and uniform, as compared to the typical QCD scales. It is nevertheless instructive to analyze the case of a uniform oscillating field, $B(t)= B_0 \cos \ov{\omega} t$, which leads to a  time-dependent vacuum polarization:
\be
\tilde{p}_z(t) &=& \hat{z}\,\theta\,\alpha^2~\frac{ 5 B_0^3 }{6  m_\eta^2~(4 \pi f_\pi)^2}\nonumber\\
&\times&\bigg\{ \bigg[ \frac{1}{6}+\frac{\overline{\omega}^2}{m_\eta^2}+ \frac{1}{10}\frac{\overline{\omega}^2}{m_\pi^2}\bigg] \cos( 3\overline{\omega} t)\nonumber\\
 &+&  \bigg[ \frac{1}{2}-\frac{\overline{\omega}^2}{3 m_\eta^2}+ \frac{11}{45}\frac{\overline{\omega}^2}{m_\pi^2}\bigg] \cos( \overline{\omega} t)\bigg\},
\label{Eq1}
\ee
where $\alpha\simeq1/137$ is the electro-magnetic fine structure constant and we have chosen a frame in which the versor $\hat{z}$ is aligned with the external magnetic field.
Eq. (\ref{Eq1}) holds in the limit $B_0\ll m_\pi^2$ and for $\ov{\omega}\ll m_\eta$. 

Interestingly, we find that the induced Vacuum Electric Dipole Moment (VEDM) displays two characteristic modes of oscillation, with frequencies $\ov{\omega}$ and $3\,\ov{\omega}$.
The origin of such modes is connected with the fact that, to lowest order in our chiral counting, the vacuum fluctuations interact with three external electro-magnetic field lines (see Fig.\ref{1Loop}). 
The mode with frequency $3\,\ov{\omega}$ corresponds to processes in which virtual states in vacuum fluctuations absorb an energy quantum $\ov{\omega}$ from each of the three external  lines. On the other hand, the mode with frequency $\ov{\omega}$ corresponds to events in which virtual states release one quantum of energy to one or two external field lines.  
We expect higher order electro-magnetic interactions to give raise  to a numerable infinity of additional characteristic oscillation modes, with strengths suppressed by higher powers of fine structure constant $\alpha$.

The induced VEDM in the presence of an static external field is readily obtained from (\ref{Eq1}) by taking the limit $\overline{\omega}~\to~0$ and reads
\be
\tilde{p}_z(t) &=& \hat{z}\,\theta\,\alpha^2~\frac{ 5 B_0^3 }{9  m_\eta^2~(4 \pi f_\pi)^2}.
\label{polT0}
\ee
It should be stressed that this formula has been obtained in a model-independent and parameter-free way. On the other hand,  it corresponds  to the lowest-order in $\theta$ and in the combined chiral and $1/N_c$ expansion.

The microscopic dynamical mechanism underlying the anomalous CP-odd electric polarization has been explored using an instanton liquid model in~\cite{faccioli}. It was shown that the instanton-mediated correlations provided by the $\theta$-term lead to a flavor- and spin-dependent repulsion between quarks and antiquarks. As a result of such an interaction, $u(d)$ quarks (antiquarks) are pushed in the direction parallel (anti-parallel) to their spin. The external magnetic field considered here generates a polarization of the quark and antiquark magnetic moments and therefore selects the direction for the electric polarization of vacuum quantum fluctuations. 

Let us now discuss the corrections to Eq.(\ref{polT0}) which arise at finite temperature. 
Since we are relying on a low-energy effective description, in the present work we can only consider temperatures much below the de-confinement temperature, $T\ll\,T_c\simeq~160$~MeV. At low temperatures, the heat-bath consists primarily of soft pions and we can use the approximation
\be
\langle J_0 \rangle_T \simeq \langle J_0 \rangle_0 + \sum_{n} \int \frac{d^3 {\bf p}}{(2\pi)^3 2\omega_\pi({\bf p})}~
\frac{\langle \pi_n({\bf p}) |J_0| \pi_n({\bf p}) \rangle}{e^{\frac{\omega_\pi({\bf p})}{T}}-1 }.\nonumber
\ee
We find that, for a static magnetic field  along the $\hat{z}$ axis,  the leading chiral-order expression for
the induced electric polarization is 
\be
p_z\simeq \hat{z} \frac{5~ \theta\,\alpha\,B_0}{9 (4 \pi f_\pi)^2~m_\eta^2} \left( \alpha B_0^2 + \frac{5 \pi ~m_\pi^2}{16} ~T^2+\cdots\right)
\label{IEDMT}
\ee
We note that the finite temperature correction displays  qualitative differences with respect to the $T=0$ contribution: it is of order $\alpha$ (rather that $\alpha^2$), and grows  linearly (rather than cubically) with the external magnetic field. 
From Eq.(\ref{IEDMT}) it follows that the energy transferred to a charged particle propagating through the uniformly polarized vacuum for a distance $L$ is proportional to $\alpha^2 \theta B^3 L$, with  temperature corrections of the order ~$\alpha\theta B L T^2$.

Let us now discuss some of the phenomenological implications of these results.  We first analyze  the magnitude of the polarization induced on cosmological distance scales by the LSMF.
It is immediate to realize that the $T=0$ contribution is extremely small. To see this, we consider a box of side size 
$L= 10^2$~kpc, permeated by a field of 1 $\mu G=10^{-10}$~T. This  represents a typical configuration for an observed domain permeated by a LSMF. We find that the induced electric potential difference at the opposite sides of the box would be only $\Delta V \simeq \theta 10^{-35}$~V. 
Given the smallness of this effect, it is extremely unlikely that present LSMF could provide observable signatures of strong CP violation. 

The situation is partially modified at finite temperature. We consider a scenario in which $T\sim150$~MeV, $B_0\sim0.1~$nG and the magnetic domain is modeled as a box of two astronomical unit side. This configuration represents the most favorable compatible with the primordial hypothesis for the LSMF, at the end of the hadron epoch. We find that the induced VEDM would be $D\sim 2\times 10^{31}\theta$ e\,cm  --- i.e. about $10^{45}$ times the electric dipole moment of a single neutron---, corresponding to an induced surface pion density of $n\sim 2\times10^{-4} \theta \textrm{m}^{-2}$  and a potential energy difference at the side of the box of $\Delta V\sim \theta \times 1~$V.
This result shows that, if $\theta\ne 0$ during the hadron epoch, then  very small  finite separation of electric (and flavor) charge was  induced in the regions permeated by the static primordial fields. 

We stress the fact that, although the effect is still extremely small, we have found that it is strongly enhanced at finite temperature. Hence, it would be interesting to investigate the magnitude of the polarization at much higher temperatures, which were reached in earlier stages of the Big Bang. Above the QCD critical temperature $T_c\simeq 160~$MeV quarks and gluons are de-confined and the low-energy effective description adopted here brakes down. However, the dominant topological correlations in the large temperature limit can systematically be accounted for by computing the effect of  perturbative fluctuations around the small-sized caloron configurations.

In passing, we note that a charge asymmetry in pion momentum distribution has also been discussed in the context of ultra-relativistic heavy-ion collisions~\cite{CPRICH}, assuming the formation of meta-stable CP-odd "false vacua", where $\theta\sim1$. In heavy ion collisions the direction of the asymmetry is provided by the total orbital angular momentum, which is perpendicular to the collision plane. 

Let us now estimate the magnitude of the induced VEDM if the magnetic field is extremely intense. Such a scenario is realized in near the accretion disk of a black hole or in the vicinity of  magnetar, where the magnetic field strength can be as high as $10^{15}\div 10^{16}$~G.
A straightforward calculation from Eq. (\ref{polT0}) shows that the energy transfered by the VEDM to a charge traveling for $10^{4}$~m --- i.e. a distance of the order of the radius of a neutron star--- is  $\Delta E \sim~\theta~\times 1~$TeV. 
Choosing $B_0\sim 10^{15}$~G and $\theta\sim 10^{-10}$ ---i.e. the largest value compatible with the measurement of the neutron electric dipole moment--- one finds that electrons can be accelerated to about $0.1$~KeV. This implies that the present CP-odd effect may in principle play a role in processes related to photon emissions from compact objects up frequency in the $X$-ray range. 

We stress the fact that the value $B\sim 10^{15}$G is a rather conservative one. If the magnetic field strength is chosen to be just one order of magnitude larger (see e.g. the discussion in~\cite{magnetar}) then the energy transferred to electrons would be increased by three orders of magnitude and enter the regime relevant for $\gamma-$ray emission.  

In view of such a result, it is interesting to take a closer look to the induced VEDM generated by the magnetic field of a magnetar. If  we neglect the effects associated to the rotations of the star, then the magnetic field can be assumed to be  purely dipolar~\cite{pulsar}. In this case, the resulting total  induced electric dipole moment can be easily computed by performing the integral Eq.(\ref{edm0}) over the entire space. For a typical magnetic moment $m=10^{27} \textrm{G m}^3$, we find an induced electric dipole moment of $D\sim 3\times 10^{26} \theta$~e\,cm.
It should be noted  that, if the star  rotates with angular velocity ${\bf \Omega}$ directed along the axis parallel to its magnetic moment,  an additional CP-even electric field is developed~\cite{pulsar}. However, the resulting charge polarization would lead to qualitatively different electric fields. In fact, it can be shown that in this case  the induced charge density  is in the form $\rho(x) = -1/(2\pi)\,{\bf \Omega \cdot B}$, hence it only  contributes to the electric quadrupole moment of the star. 

A detailed investigation of the phenomenological consequences of the CP-odd electric dipole moment of a magnetar is beyond the scope of the present work. Here, we only point out that the anomalous electric currents generated by the VEDM would be most intense near the poles of the star, where they would induce a local enhancement of the temperature. This phenomenon may have astrophysical implications: in fact, the existence of "hot spots" near the surface of neutron stars has been suggested as a possible scheme to explain the structure of their spectrum  of emission  below the $X-$ray bend, which cannot be fitted by a single black-body radiation formula ~\cite{spec}. 

In conclusion, in this work we have studied the response of the QCD vacuum to an external magnetic field, in the presence of the strong CP-odd correlations introduced by the $\theta$-term. We have shown that an external magnetic field would induce a separation of electric and flavor charge, associated to the polarization of quantum vacuum fluctuations. A parameter free formula for the induced polarization has been obtained in the framework of chiral perturbation theory and large $N_c$ expansion.

We have estimated the magnitude of this effect by considering the field generated in the presence of LSMF and by the strong magnetic fields present near neutron stars or black holes.
We have found that the polarization induced by the observed LSMF is extremely small and we do not expect it to have phenomenological implications. Nevertheless, at least at the temperature regime accessible to our analysis, the polarization increases significantly with $T$.  On the basis of such a result, we believe that it is important to compute the induced polarization at temperatures much above the QCD phase transition, where the effect could be strongly enhanced. 

The polarization induced by the magnetic field in the vicinity of neutron stars can accelerate electrons up to energies of the order $\theta \times$~TeV. Using the present upper bound on the value of $\theta$ one finds that in principle the VEDM may play a role in processes related to photon emissions from compact objects up to  frequencies in the $X$-ray range ---for fields of the order of $10^{15}$~G--- or even $\gamma$-ray ---for fields of the order of $10^{16}~G$---. 
We have also suggested a connection with the possible existence of "hot-spots" on the surface of neutron stars.

  
This work was motivated by a discussion with A.~Zhitnitsky. We thank V. Pascalutsa, A. Steiner, D.Blaschke and E.V. Shuryak for important discussions. Feynman diagrams have been drawn using Jaxodraw\cite{jaxodraw}.

\end{document}